\documentclass[preprint,12pt]{elsarticle}

\usepackage{amssymb}
\usepackage{amsmath}
\usepackage{booktabs}
\usepackage{float}
\usepackage{adjustbox}
\usepackage{xcolor}
\usepackage{soul}

\journal{Journal of Power Sources}

\begin{document}

\begin{frontmatter}

%% Title, authors and addresses

%% use the tnoteref command within \title for footnotes;
%% use the tnotetext command for theassociated footnote;
%% use the fnref command within \author or \address for footnotes;
%% use the fntext command for theassociated footnote;
%% use the corref command within \author for corresponding author footnotes;
%% use the cortext command for theassociated footnote;
%% use the ead command for the email address,
%% and the form \ead[url] for the home page:
%% \title{Title\tnoteref{label1}}
%% \tnotetext[label1]{}
%% \author{Name\corref{cor1}\fnref{label2}}
%% \ead{email address}
%% \ead[url]{home page}
%% \fntext[label2]{}
%% \cortext[cor1]{}
%% \affiliation{organization={},
%%             addressline={},
%%             city={},
%%             postcode={},
%%             state={},
%%             country={}}
%% \fntext[label3]{}

\title{A Control-Oriented Simplified Single Particle Model with Grouped Parameter and Sensitivity Analysis for Lithium-Ion Batteries}

%% use optional labels to link authors explicitly to addresses:
%% \author[label1,label2]{}
%% \affiliation[label1]{organization={},
%%             addressline={},
%%             city={},
%%             postcode={},
%%             state={},
%%             country={}}
%%
%% \affiliation[label2]{organization={},
%%             addressline={},
%%             city={},
%%             postcode={},
%%             state={},
%%             country={}}

\author[inst1,inst2]{Feng Guo \corref{cor1}}
\author[inst1,inst2]{Luis D. Couto}

\affiliation[inst1]{organization={VITO},%Department and Organization
            addressline={Boeretang 200}, 
            city={Mol},
            postcode={2400}, 
            country={Belgium}}

\affiliation[inst2]{organization={EnergyVille},%Department and Organization
            addressline={Thor Park 8310}, 
            city={Genk},
            postcode={3600}, 
            country={Belgium}}

\cortext[cor1]{Corresponding author:feng.guo@vito.be(Feng Guo)}

\begin{abstract}
%% Text of abstract
Lithium-ion batteries are widely used in transportation, energy storage, and consumer electronics, driving the need for reliable battery management systems (BMS) for state estimation and control. The Single Particle Model (SPM) balances computational efficiency and accuracy but faces challenges in parameter estimation due to numerous parameters. Current SPM models using parabolic approximation introduce intermediate variables and hard to do parameter grouping. This study presents a control-oriented SPM reformulation that employs parameter grouping and parabolic approximation to simplify model parameters while using average and surface lithium-ion concentrations as model output. By parameter grouping, the original 17 parameters were reduced to 9 grouped parameters. The reformulated model achieves a reduced-order ordinary differential equation form while maintaining mathematical accuracy equivalent to the pre-grouped discretized SPM. Through Sobol sensitivity analysis under various current profiles, the grouped parameters were reduced from 9 to 6 highly sensitive parameters. Results demonstrate that estimating these 6 parameters achieves comparable practical accuracy to estimating all 9 parameters, with faster convergence. This control-oriented SPM enhances BMS applications by facilitating state estimation and control while reducing parameter estimation requirements.

\end{abstract}

%%Graphical abstract
% \begin{graphicalabstract}
% \includegraphics[width=1\textwidth]{}
% \end{graphicalabstract}

% %%Research highlights
% \begin{highlights}
% \item SPM discretized using parabolic approximation and grouped into an ODE form.
% \item Global sensitivity analysis applied to reduce parameters for estimation.
% \item Reduced 17 parameters to 6 using sensitivity analysis and parameter grouping.
% \item Faster convergence achieved by estimating only highly sensitive parameters.

% \end{highlights}

\begin{keyword}
%% keywords here, in the form: keyword \sep keyword
Lithium-ion batteries \sep Electrochemical Model \sep Control-Oriented \sep Parameter Estimation \sep Parameter Grouping \sep Sensitivity Analysis

\end{keyword}

\end{frontmatter}

%% \linenumbers

%% main text
\section{Introduction}
\label{sec:intro}
With the increasing use of lithium-ion batteries in transportation applications, energy storage, and consumer electronic\cite{peng2024enhancing}, researchers and industry professionals are becoming more concerned with battery safety and optimizing battery performance \cite{jiang2024advances, nyamathulla2024review}. Physically-based electrochemical models are increasingly being used for battery aging \cite{reniers2019review}, state estimation \cite{guo2024systematic,li2021adaptive}, and control algorithms \cite{couto2021faster}. However, these models are complex and present numerous challenges in practical applications, particularly in parameter estimation \cite{andersson2022parametrization,rojas2024critical}. Typically, specialized experimental equipment is required to measure certain parameters, and some cannot be directly measured, and must be identified through parameter estimation methods. Additionally, electrochemical models are often over-parameterized, meaning that different sets of parameters can yield the same model outcomes, which further complicates parameter identification \cite{miguel2021review}.

Electrochemical models commonly refer to pseudo-two-dimensional (P2D) models and their simplified versions. The P2D model is based on the microscopic electrochemical reaction mechanisms within the battery \cite{fuller1994simulation}. However, due to its complexity and high computational demands, a simplified version called the Single Particle Model (SPM) is often used in practical applications. The SPM replaces the multiple particles in the P2D model with a single averaged particle and substitutes the complex electrolyte reactions with a fixed average concentration. By incorporating the reactions in the electrolyte into the SPM, the SPM with electrolyte (SPMe) can be obtained \cite{marquis2019asymptotic}. This simplification makes the SPM and SPMe more straightforward and significantly faster in computation compared to the P2D model \cite{haran1998determination}. In real-world control system applications, considering the computational capabilities of microprocessors in battery management system (BMS), it is often necessary to simplify the model to reduce the number of parameters and computational burden. The SPM is often the optimal choice in such scenarios. The most commonly used methods for parameter estimation for electrochemical models are the Least Squares method \cite{zheng2016co}, Particle Swarm Optimization (PSO) \cite{fan2020systematic,shao2023novel}, and Genetic Algorithms (GA) \cite{feng2020co}. The least squares method has the advantage of being very fast, but it tends to converge to local optima. On the other hand, PSO and GA methods are more capable of finding the global optimum, but they require longer computation times. Additionally, due to the numerous parameters in electrochemical models, it is usually necessary to group them to reduce the estimation workload.

Parameter grouping ensures that the reformulated model maintains the same mathematical accuracy as the original model, while reducing the number of parameters requiring identification. The core idea of parameter grouping is to combine independent parameters into groups, so that during parameter estimation, only the grouped parameters are estimated, rather than estimating each parameter individually. For example, consider the equation 
\(y = (A \times B)x + C,\) 
where \(x\) is the input, \(y\) is the output, and \(A\), \(B\), and \(C\) are parameters. By defining \(S = A \times B,\) we can estimate \(S\) instead of attempting to estimate \(A\) and \(B\) separately, as \(A\) and \(B\) can have infinitely many combinations that yield the same result. Similarly, in electrochemical models, many parameters are like \(A\) and \(B\) and cannot be estimated independently. Directly estimating each of these parameters would be difficult, making parameter grouping a practical approach \cite{miguel2021review}. Drummond et al. conducted the parameter grouping for a P2D model, determining that the model can be fully characterized by estimating 21 parameters \cite{drummond2020structural}. Chu et al. also performed parameter grouping on a P2D model, ultimately reducing the number of parameters from 36 to 24 \cite{chu2019control}. Namor et al. proposed a parameter grouping method for the SPM, reducing the number of parameters to seven \cite{namor2017parameter}. Subsequently, Bizeray et al. also conducted parameter grouping for an SPM, concluding that only six parameters are necessary for accurate model identification \cite{bizeray2018identifiability}. The additional parameter in Namor et al.'s model compared to Bizeray et al.'s is due to the inclusion of the battery's internal resistance in Namor et al.'s SPM. It is important to note that the above parameter sets do not account for the initial state of charge (SOC) of the positive and negative electrodes. These two parameters must also be estimated during parameter identification, thereby increasing the total number of parameters to be estimated by two.

Parameter sensitivity analysis is used to evaluate how changes in model parameters affect the model's output \cite{ye2017global}. This analysis is typically divided into two types: local sensitivity analysis and global sensitivity analysis \cite{andersson2022parametrization,rojas2024critical}. 
Local sensitivity analysis involves changing one parameter at a time while keeping all other parameters constant, to assess its impact on the model output. Common methods for local sensitivity analysis include derivative-based and variance-based methods. In derivative-based methods, the sensitivity of a parameter is determined by calculating the partial derivative of the model output with respect to that parameter; the magnitude of the derivative indicates the sensitivity \cite{park2018optimal,jin2018parameter,lai2020optimization}. Variance-based methods, on the other hand, assess how changes in a single parameter within a specified range affect the model output, typically using standard deviation as a measure of sensitivity \cite{edouard2016parameter,li2020parameter,liu2020simulation}. Local parameter sensitivity analysis is limited because it changes only one parameter at a time, exploring a limited portion of the parameter space and not accounting for interactions between parameters. This approach is therefore often considered limited in scope. 

In contrast, global sensitivity analysis involves varying all parameters simultaneously, allowing for exploration of the entire parameter space. As a result, global sensitivity analysis provides more comprehensive and valuable insights compared to local sensitivity analysis \cite{saltelli2019so}. Commonly used global sensitivity analysis methods include Analysis of Variance (ANOVA) \cite{vazquez2014rapid}, the Morris Method \cite{grandjean2019global}, and the Sobol Method \cite{fan2021global,streb2023investigating,laue2020model}.Vazquez-Arenas et al. conducted a sensitivity analysis of the parameters in the P2D model using ANOVA \cite{vazquez2014rapid}.  Grandjean et al. conducted a global sensitivity analysis of the SPMe's parameters using the Morris Method \cite{grandjean2019global}. Fan et al. employed the Sobol method to conduct a global sensitivity analysis of battery parameters \cite{fan2021global}. ANOVA results are presented in terms of variance components, providing the main effects of factors on the output as well as second-order interaction effects. This method is particularly suitable when the number of input factors is small and the model structure is relatively simple \cite{montgomery2017design}. The Morris Method results, summarized by mean and standard deviation, provide a general ranking of parameter influence, highlighting overall effects and non-linear interactions \cite{petropoulos2016sensitivity}. In contrast, the Sobol method offers detailed sensitivity indices, including first-order, higher-order, and total effects, giving a comprehensive analysis of parameter contributions \cite{campolongo2007effective}. In summary, if accurate global sensitivity analysis results are desired, the Sobol method is a good choice. 

Specifically, when dealing with battery parameter grouping, partial differential equations (PDE) in electrochemical models must be discretized using methods such as the parabolic approximation method \cite{subramanian2005efficient} to convert them into ordinary differential equations (ODE) for practical applications. However, this process often introduces new variables or alters parameter combinations, complicating parameter grouping. Namor et al. \cite{namor2017parameter}, while analyzing battery parameter grouping, did not consider the impact of these discretization methods. In contrast, Bizeray et al. \cite{bizeray2018identifiability} transformed the electrochemical model into the frequency domain using transfer functions, but this required additional Electrochemical Impedance Spectroscopy (EIS) data, adding complexity to the parameter identification process. These challenges highlight the need for parameter grouping approaches that consider the impact of discretization while maintaining model accuracy and practicality for real-world applications.

In this context, parameter sensitivity analysis has emerged as a complementary approach to improve model efficiency. Most research has primarily focused on presenting sensitivity analysis results, with limited application of these results to enhance the model or parameter estimation process. For instance, Zhang et al. \cite{zhang2014parameter} and Park et al. \cite{park2018optimal} used sensitivity analysis to guide parameter estimation by prioritizing highly sensitive parameters. However, their approach required multiple iterations, making it time-intensive. Khalik et al. \cite{khalik2021parameter} performed a parameter sensitivity analysis on the P2D model. Based on the results, they ranked the parameters by sensitivity and conducted multiple parameter estimation simulations to determine the optimal number of aggregated parameters required, thereby reducing the number of parameters that needed to be estimated. However, the process was also quite time-consuming. Guo et al. \cite{guo2019equivalent} conducted a sensitivity analysis on equivalent circuit models and demonstrated that estimating only highly sensitive parameters could achieve comparable accuracy to estimating all parameters, with significantly improved computational speed. Building on these findings, integrating parameter sensitivity information into the parameter grouping process for electrochemical models offers a promising direction to streamline parameter estimation while maintaining model accuracy and efficiency.

Based on the above discussion, the main contributions of this paper are as follows:

\begin{enumerate}
    \item A control-oriented reformulation of the SPM is proposed by discretizing it using a parabolic approximation method. Grouping parameters leads to an ODE-based model that retains mathematical accuracy while reducing parameter complexity, making it more suitable for state estimation and control applications.
    \item A global sensitivity analysis of the grouped-parameter SPM is performed using the Sobol method across various constant and dynamic current profiles, identifying highly sensitive parameters critical for control and further reducing the parameters required for estimation.

\end{enumerate}

\section{SPM}
\label{sec:SPM}

This study is based on the SPM.
The governing equation for lithium concentration in the solid phase is given by:
\begin{equation}
\frac{\partial c_{s,i}}{\partial t}(r,t) = \frac{D_{s,i}}{R_{s,i}^2} \frac{\partial}{\partial r} \left( r^2 \frac{\partial c_{s,i}}{\partial r}(r,t) \right),\label{eq:1}
\end{equation}
where \( i \in \{p,n\} \) corresponds to the positive and negative electrodes, respectively, \( c_{s} \) is the lithium concentration in the solid phase, \( D_{s} \) is the solid-phase diffusion coefficient, \( R_{s} \) is the particle radius, \( r \) is the radial position within the electrode particle, and \( t \) is time. 
The initial conditions can be defined as uniform like $c_{s,i}(r,0) = c_i(0)$.

The boundary conditions for the PDE Eq.\eqref{eq:1} are specified as follows: at the center of the electrode particle,
\begin{equation}
\left. \frac{\partial c_{s,i}}{\partial r}(r,t) \right\vert_{r=0} = 0,\label{eq:2}
\end{equation}
and at the particle's surface,
\begin{equation}
\left. \frac{\partial c_{s,i}}{\partial r}(r,t) \right\vert_{r=R_{s,i}} = \frac{-j_i (t)}{D_{s,i}},  \label{eq:3}
\end{equation}
where \( j_i \) is the flux of lithium ions into the particle.

The terminal cell voltage \( V_{{\rm SPM}} \) is computed as:
\begin{equation}
\begin{aligned}
V_{{\rm SPM}} (t) = OCP_p(\tilde{c}_{ss,p}(t)) - OCP_n(\tilde{c}_{ss,n}(t)) + \\
\eta_{p}(\tilde{c}_{ss,p}(t), I(t)) - \eta_{n}(\tilde{c}_{ss,n}(t), I(t)) - R_0 I(t), 
\end{aligned} 
\label{eq:4}
\end{equation}
where \( OCP(\tilde{c}_{ss,i}(t)) \) represents the open-circuit potential, the detailed formulations for the positive and negative electrodes can be found in Appendix A. \( \eta_{i}(\tilde{c}_{ss,i}(t),I(t)) \) is the surface overpotential, \( I \) is the applied current, and \( R_0 \) is the internal resistance of the cell. The  normalized surface concentration \( \tilde{c}_{ss,i}(t) \) is defined as:
\begin{equation}
\tilde{c}_{ss,i}(t) = \frac{c_{ss,i}(t)}{c_{{\rm max},i}}, \label{eq:5}
\end{equation}
where \( c_{ss} \) is the solid-phase lithium concentration at the surface of the spherical particle (i.e. $c_{s,i}(r,t)\vert_{r = R_{s,i}} = c_{ss,i}(t)$), and \( c_{{\rm max}} \) is the maximum solid-phase lithium concentration.

The surface overpotential \( \eta_{i}(c_{ss,i}(t), I(t)) \) is given by:
\begin{equation}
\eta_{i}(\tilde{c}_{ss,i}(t), I(t)) = \frac{2RT}{F} \sinh^{-1} \left( \frac{1_\mp I(t)}{2a_i L_i j_{0,i}(\tilde{c}_{ss,i}(t))} \right), \label{eq:6}
\end{equation}
where \( 1_\mp \) is a sign indicator with \( -1 \) and \( +1 \) for positive and negative electrode, respectively,
\( R \) is the universal gas constant, \( T \) is the temperature, and \( F \) is Faraday's constant. The variables \( a \) and \( L \) represent the specific surface area and thickness of the electrode, respectively. The exchange current density \( j_{0,i}(\tilde{c}_{ss,i}(t)) \) is expressed as:
\begin{equation}
j_{0,i}(\tilde{c}_{ss,i}(t)) = r_{{\rm eef},i} c_{{\rm max},i} \sqrt{c_e \tilde{c}_{ss,i}(t) (1 - \tilde{c}_{ss,i}(t))}, \label{eq:7}
\end{equation}
where \( r_{{\rm eef}} \) is the electrode reaction rate constant, and \( c_e \) is the lithium-ion concentration in the electrolyte.

The intercalation current density \( j_{i}(t) \) is assumed to be uniform along the cell thickness, leading to the following representation:
\begin{equation}
j_{i}(t) = %\pm 
1_\mp \frac{I(t)}{Fa_iA_iL_i}, \label{eq:8}
\end{equation}
where \( A_i \) is the electrode surface area. 

Among the parameters listed in the above model, \( R \) and \( F \) are constants, meaning they do not require estimation. The parameter \( c_e \) is typically set as a constant in the SPM, often specified as 1000 \({\rm mol}\cdot{\rm m}^{-3}\), and therefore also does not require estimation. Additionally, the specific surface area \( a_i \) can be calculated from other parameters, so it does not need to be estimated independently:
\begin{equation}
a_i = \frac{3 \varepsilon_{s,i}}{R_{s,i}}, \label{eq:9}
\end{equation}
It is also worth noting that the initial concentrations of lithium in the solid phase  for both the positive (\(c_{p}(0)\)) and negative (\(c_{n}(0)\)) electrodes usually need to be estimated. Therefore, the final set of parameters requiring estimation consists of 17 parameters in total:

\begin{equation}
\begin{aligned}
P_{\text{all}} = \big[ &D_{s,n}, D_{s,p}, R_{s,n}, R_{s,p}, r_{\text{eef},n}, r_{\text{eef},p}, \\
                             &A_n, A_p, L_n, L_p, c_{\text{max},n}, c_{\text{max},p}, \\
                             &\varepsilon_{s,n}, \varepsilon_{s,p},c_{n}(0), c_{p}(0) , R_0 \big],
\end{aligned}
\label{eq:10}
\end{equation}

\section{Model Discretization and Parameter Grouping}
\label{sec:MDPG}

This section we presents the discretization of the SPM using the parabolic approximation method and the subsequent grouping of model parameters. The parabolic approximation method employs a parabola to approximate the diffusion process described in Eq. \eqref{eq:1}, transforming it into an ODE for easier computation \cite{subramanian2005efficient}. The governing equations for the parabolic approximation are given as follows \cite{tagade2016recursive}:

\begin{equation}
\frac{d}{dt} \overline{c}_{s,i} = -\frac{3}{R_{s,i}}j_i
\label{eq:11}
\end{equation}

\begin{equation}
\frac{d}{dt} \overline{c}_{fs,i} + 30\frac{D_{s,i}}{R_{s,i}^{2}} \overline{c}_{fs,i} + \frac{45}{2R_{s,i}^{2}} j_i = 0 
\label{eq:12}
\end{equation}

\begin{equation}
c_{ss,i} = \overline{c}_{s,i} + \frac{8R_{s,i}}{35}\overline{c}_{fs,i} - \frac{R_{s,i}}{35D_{s,i}}j_i 
\label{eq:13}
\end{equation}
where $\overline{c}_{fs}$ is the average concentration flux of lithium in the active material. From the above equation, the intermediate variable $\overline{c}_{fs}$ has no direct physical significance, the introduction of variable $\overline{c}_{fs}$ makes parameter grouping challenging.

Therefore, in the following sections, we reformulate the model by using the average lithium concentration $\overline{c}_{s,i}$ and surface lithium concentration 
 $c_{ss,i}$ as output and perform parameter grouping.

To further develop the model, consider the following state-space representation of the parabolic approximation model:
\begin{align}
    \label{eq:14}
    \dot{x}(t) & = A x(t) + B u(t), \quad x(0) = c_0 \\
    \label{eq:15}
    y(t) & = C x(t) + D u(t),
\end{align}
where the input \( u = I(t) \) represents the applied current, and the state vector \( x \) is given by
\begin{equation}
x = \begin{bmatrix} x_p^\top & x_n^\top \end{bmatrix}^\top,
\label{eq:16}
\end{equation}
with \( x_p \) and \( x_n \) representing the state vectors of the positive and negative electrodes, respectively. Each state vector is defined as
\begin{equation}
x_i = \begin{bmatrix} \overline{c}_{s,i} & \overline{c}_{f,i} \end{bmatrix}^\top,
\label{eq:17}
\end{equation}
where \( i \in \{p, n\} \). The initial conditions are
\begin{equation}
x_i(0) = \begin{bmatrix} \overline{c}_{s,i}(0) & \overline{c}_{f,i}(0) \end{bmatrix}^\top.
\label{eq:18}
\end{equation}
The output vector \( y \) is defined as
\begin{equation}
y = \begin{bmatrix} y_p^\top & y_n^\top \end{bmatrix}^\top,
\label{eq:19}
\end{equation}
where
\begin{equation}
y_i = \begin{bmatrix} \overline{c}_{s,i} & c_{ss,i} \end{bmatrix}^\top,
\label{eq:20}
\end{equation}

The state-space matrices \( A = \text{diag}(A_p, A_n) \) and \( B = \begin{bmatrix} B_p^\top & B_n^\top \end{bmatrix}^\top \) are defined as:
\begin{equation}
A_i = \begin{bmatrix}
0 & 0 \\ 
0 & \displaystyle -\frac{30 D_{s,i}}{R_{s,i}^2}
\end{bmatrix}, \quad
B_i = \begin{bmatrix}
\displaystyle \frac{1}{F A_i L_i \varepsilon_i} \\
\displaystyle \frac{15}{2 R_{s,i}} \frac{1}{F A_i L_i \varepsilon_i}
\end{bmatrix}.
\label{eq:21}
\end{equation}

Similarly, the output matrices \( C = \text{diag}(C_p, C_n) \) and \( D = \begin{bmatrix} D_p^\top & D_n^\top \end{bmatrix}^\top \) are given by:
\begin{equation}
C_i = \begin{bmatrix}
1 & 0 \\ 
1 & \displaystyle \frac{8 R_{s,i}}{35}
\end{bmatrix}, \quad
D_i = \begin{bmatrix}
0 \\
\displaystyle \frac{R_{s,i}^2}{105 D_{s,i}} \frac{1}{F A_i L_i \varepsilon_i}
\end{bmatrix}.
\label{eq:22}
\end{equation}

From the equations above, it is evident that parameter grouping is challenging, particularly because the \( C_i \) matrix contains an isolated \( R_{s,i} \) parameter that cannot be easily integrated with other parameters. This difficulty arises due to the introduction of the variable \( \overline{c}_{f,i} \) during the parabolic approximation discretization process. To address this, we rewrite the state-space equations by directly incorporating the surface lithium-ion concentration (\( c_{ss,i} \)) as a state variable, thereby eliminating the variable \( \overline{c}_{f,i} \).
To facilitate parameter grouping, consider the following transformation of variables:
\begin{equation}
c_{ss,i} = \overline{c}_{s,i} + \frac{8}{35} R_{s,i} \overline{c}_{f,i} + \frac{R_{s,i}^2}{105 D_{s,i}} \frac{1}{F A_i L_i \varepsilon_i} I(t),
\label{eq:23}
\end{equation}
which allows the state-space model to be rewritten as:
\begin{align}
    \label{eq:24}
    \begin{bmatrix}
    \dot{\overline{c}}_{s,i}(t) \\ 
    \dot{c}_{ss,i}(t)
    \end{bmatrix} 
    & = 
    \begin{bmatrix}
    \displaystyle \frac{1}{F A_i L_i \varepsilon_i} I(t) \\ 
    \displaystyle \frac{30 D_{s,i}}{R_{s,i}^2} \left(\overline{c}_{s,i}(t) - c_{ss,i}(t)\right) + \frac{3}{F A_i L_i \varepsilon_i} I(t) + \frac{R_{s,i}^2}{105 D_{s,i}} \frac{1}{F A_i L_i \varepsilon_i} \dot{I}(t)
    \end{bmatrix}.
\end{align}

Next, define the following grouped parameters: \( \alpha_i = {R_{s,i}^2}/{D_{s,i}} \) and \( \beta_i = F A_i L_i \varepsilon_i \). With these, the model can be further simplified:
\begin{align}
    \label{eq:25}
    \begin{bmatrix}
    \dot{\overline{c}}_{s,i}(t) \\ 
    \dot{c}_{ss,i}(t)
    \end{bmatrix} 
    & = 
    \begin{bmatrix}
    \displaystyle \frac{1}{\beta_i} I(t) \\ 
    \displaystyle \frac{30}{\alpha_i} \left(\overline{c}_{s,i}(t) - c_{ss,i}(t)\right) + \frac{3}{\beta_i} I(t) + \frac{\alpha_i}{105 \beta_i} \dot{I}(t)
    \end{bmatrix}.
\end{align}

To eliminate the derivative of the input \( \dot{I}(t) \), apply the transformation:
\begin{equation}
\label{eq:26}
\begin{bmatrix}
q_{1,i}(t) \\ 
q_{2,i}(t)
\end{bmatrix} 
= 
\begin{bmatrix}
\displaystyle \overline{c}_{s,i}(t) \\ 
\displaystyle c_{ss,i}(t) - \frac{\alpha_i}{105 \beta_i} I(t)
\end{bmatrix}.
\end{equation}

The resulting model is then given by:
\begin{align}
    \label{eq:27}
    \begin{bmatrix}
    \dot{q}_{1,i}(t) \\ 
    \dot{q}_{2,i}(t)
    \end{bmatrix} 
    & = 
    \begin{bmatrix}
    \displaystyle \frac{1}{\beta_i} I(t) \\ 
    \displaystyle \frac{30}{\alpha_i} \left(q_{1,i}(t) - q_{2,i}(t)\right) + \frac{19}{7 \beta_i} I(t)
    \end{bmatrix},
\end{align}
with output variables:
\begin{align}
    \label{eq:28}
    \begin{bmatrix}
    \overline{c}_{s,i}(t) \\ 
    c_{ss,i}(t)
    \end{bmatrix} 
    & = 
    \begin{bmatrix}
    \displaystyle q_{1,i}(t) \\ 
    \displaystyle q_{2,i}(t) + \frac{\alpha_i}{105 \beta_i} I(t)
    \end{bmatrix}.
\end{align}

The initial conditions must be adapted according to these transformations. Since \( \overline{c}_{s,i} = q_{1,i} \), the initial conditions for \( \overline{c}_{s,i} \) are directly equivalent to \( q_{1,i} \). The initial condition for \( q_{2,i}(0) \) is given by:
\begin{equation}
q_{2,i}(0) = \overline{c}_{s,i}(0) + \frac{8}{35} R_{s,i} \overline{c}_{f,i}(0).
\label{eq:29}
\end{equation}

Finally, the reformulated model can be expressed in the state-space form as:
\begin{align}
    \label{eq:30}
    \dot{\chi}(t) & = \tilde{A} \chi(t) + \tilde{B} u(t), \quad \chi(0) = c_{i}(0)/c_{{\rm max},i}, \\
    \label{eq:31}
    \psi(t) & = \tilde{C} \chi(t) + \tilde{D} u(t),
\end{align}
where the normalized state vector is \( \chi = \begin{bmatrix} \chi_p^\top & \chi_n^\top \end{bmatrix}^\top \), with \( \chi_i = \begin{bmatrix} \tilde{q}_{1,i} & \tilde{q}_{2,i} \end{bmatrix}^\top = \begin{bmatrix} q_{1,i}/c_{{\rm max},i} & q_{2,i}/c_{{\rm max},i} \end{bmatrix}^\top \), the input is the applied current \( u = I \), the output vector is defined as \( \psi = \begin{bmatrix} \psi_p^\top & \psi_n^\top \end{bmatrix}^\top \), where \( \psi_i = \begin{bmatrix} \tilde{\overline{c}}_{s,i} & \tilde{c}_{ss,i} \end{bmatrix}^\top = \begin{bmatrix} \overline{c}_{s,i}/c_{{\rm max},i} & c_{ss,i}/c_{{\rm max},i} \end{bmatrix}^\top \).

The state matrices \( \tilde{A} = \text{diag}(\tilde{A}_p, \tilde{A}_n) \), \( \tilde{B} = \begin{bmatrix} \tilde{B}_p^\top & \tilde{B}_n^\top \end{bmatrix}^\top \) are defined as:
\begin{equation}
\tilde{A}_i = \begin{bmatrix}
0 & 0 \\ 
\displaystyle \frac{30}{\alpha_i} & \displaystyle -\frac{30}{\alpha_i}
\end{bmatrix}, \quad
\tilde{B}_i = \begin{bmatrix}
\displaystyle \frac{1}{Q_i} \\
\displaystyle \frac{19}{7 Q_i}
\end{bmatrix},
\label{eq:32}
\end{equation}
and the output matrices \( \tilde{C} = \text{diag}(\tilde{C}_p, \tilde{C}_n) \), \( \tilde{D} = \begin{bmatrix} \tilde{D}_p^\top & \tilde{D}_n^\top \end{bmatrix}^\top \) are given by:
\begin{equation}
\tilde{C}_i = \begin{bmatrix}
1 & 0 \\ 
0 & 1
\end{bmatrix}, \quad
\tilde{D}_i = \begin{bmatrix}
\displaystyle 0 \\
\displaystyle \frac{\alpha_i}{105 Q_i}
\end{bmatrix},
\label{eq:33}
\end{equation}
for the grouped parameters defined as \( \alpha_i = {R_{s,i}^2}/{D_{s,i}} \) and \( Q_i = \beta_i c_{{\rm max},i} = F A_i L_i \varepsilon_i c_{{\rm max},i} \). The physical meaning of \( Q_i \) here represents the capacity of the positive and negative electrodes.

The observability matrix is given by:
\begin{equation}
\mathcal{O} = \begin{bmatrix} 
\tilde{C}_i \\ 
\tilde{C}_i \tilde{A}_i 
\end{bmatrix}
\label{eq:34}
\end{equation}

First, we compute:

\begin{equation}
\tilde{C}_i \tilde{A}_i = \begin{bmatrix} 
1 & 0 \\ 
0 & 1 
\end{bmatrix}
\begin{bmatrix} 
0 & 0 \\ 
\frac{30}{\alpha_i} & -\frac{30}{\alpha_i} 
\end{bmatrix}
=
\begin{bmatrix} 
0 & 0 \\ 
\frac{30}{\alpha_i} & -\frac{30}{\alpha_i} 
\end{bmatrix}
\label{eq:35}
\end{equation}

Thus, the observability matrix is:

\begin{equation}
\mathcal{O} = \begin{bmatrix} 
1 & 0 \\ 
0 & 1 \\ 
0 & 0 \\ 
\frac{30}{\alpha_i} & -\frac{30}{\alpha_i} 
\end{bmatrix}
\label{eq:36}
\end{equation}

Since $(\alpha_i,\, Q_i \neq 0)$. Computing the rank, we observe that the first two rows form an identity matrix, ensuring full observability:

\begin{equation}
\text{rank}(\mathcal{O}) = 2
\label{eq:37}
\end{equation}

Since the rank of the observability matrix equals the number of states, the system is fully observable.

The controllability matrix is given by:

\begin{equation}
\mathcal{C} = \begin{bmatrix} 
\tilde{B}_i & \tilde{A}_i \tilde{B}_i
\end{bmatrix}
\label{eq:38}
\end{equation}

First, we compute:

\begin{equation}
\tilde{A}_i \tilde{B}_i =
\begin{bmatrix} 
0 & 0 \\ 
\frac{30}{\alpha_i} & -\frac{30}{\alpha_i} 
\end{bmatrix}
\begin{bmatrix} 
\frac{1}{Q_i} \\ 
\frac{19}{7 Q_i} 
\end{bmatrix}
\label{eq:39}
\end{equation}

Performing the matrix multiplication:

\begin{equation}
\tilde{A}_i \tilde{B}_i =
\begin{bmatrix} 
0 \\ 
- \frac{360 }{7Q_i\alpha_i}
\end{bmatrix}
\label{eq:40}
\end{equation}

Thus, the controllability matrix is:

\begin{equation}
\mathcal{C} = \begin{bmatrix} 
\frac{1}{Q_i} & 0 \\ 
\frac{19}{7 Q_i} & - \frac{360 }{7Q_i\alpha_i}
\end{bmatrix}
\label{eq:41}
\end{equation}

Since $(\alpha_i,\, Q_i \neq 0)$. The first column has a nonzero entry in the first row, and the second column has a nonzero entry in the second row, the matrix is full rank:

\begin{equation}
\text{rank}(\mathcal{C}) = 2
\label{eq:42}
\end{equation}

Since the rank of the controllability matrix equals the number of states, the system is fully controllable.

Thus, the system satisfies the fundamental conditions for state estimation and optimal control design.

Next, we can also rewrite Eq. \eqref{eq:6} and Eq. \eqref{eq:7} as follows:

\begin{equation}
\eta_{i}(\tilde{c}_{ss,i}(t), I(t)) = \frac{2RT}{F} \sinh^{-1} \left( \frac{ 1_\mp I(t)}{6 Q_i d_i \sqrt {\tilde{c}_{ss,i}(t) (1 - \tilde{c}_{ss,i}(t))}} \right), \label{eq:43}
\end{equation}
where:
\begin{equation}
d_i = \frac{r_{\text{eef},i}  \sqrt{c_e}}{F R_{s,i}}.
 \label{eq:44}
\end{equation}

Through the above derivation, we have successfully performed parameter grouping for the SPM model's battery parameters. The number of parameters that need to be estimated is reduced to nine:
\begin{equation}
\begin{aligned}
P_{\text{grouping}} = \big[ \alpha_{n}, \alpha_{p}, Q_n, Q_p, d_n, d_p, {\rm SOC}_{n}(0), {\rm SOC}_{p}(0),R_0 \big]
\end{aligned}
\label{eq:45}
\end{equation}

\section{Parameter Sensitivity Analysis}
\label{sec:PSA}
\begin{figure}[H]
    \centering
    \includegraphics[width=\textwidth,trim={0.3cm 0.3cm 0.3cm 0.3cm},clip]{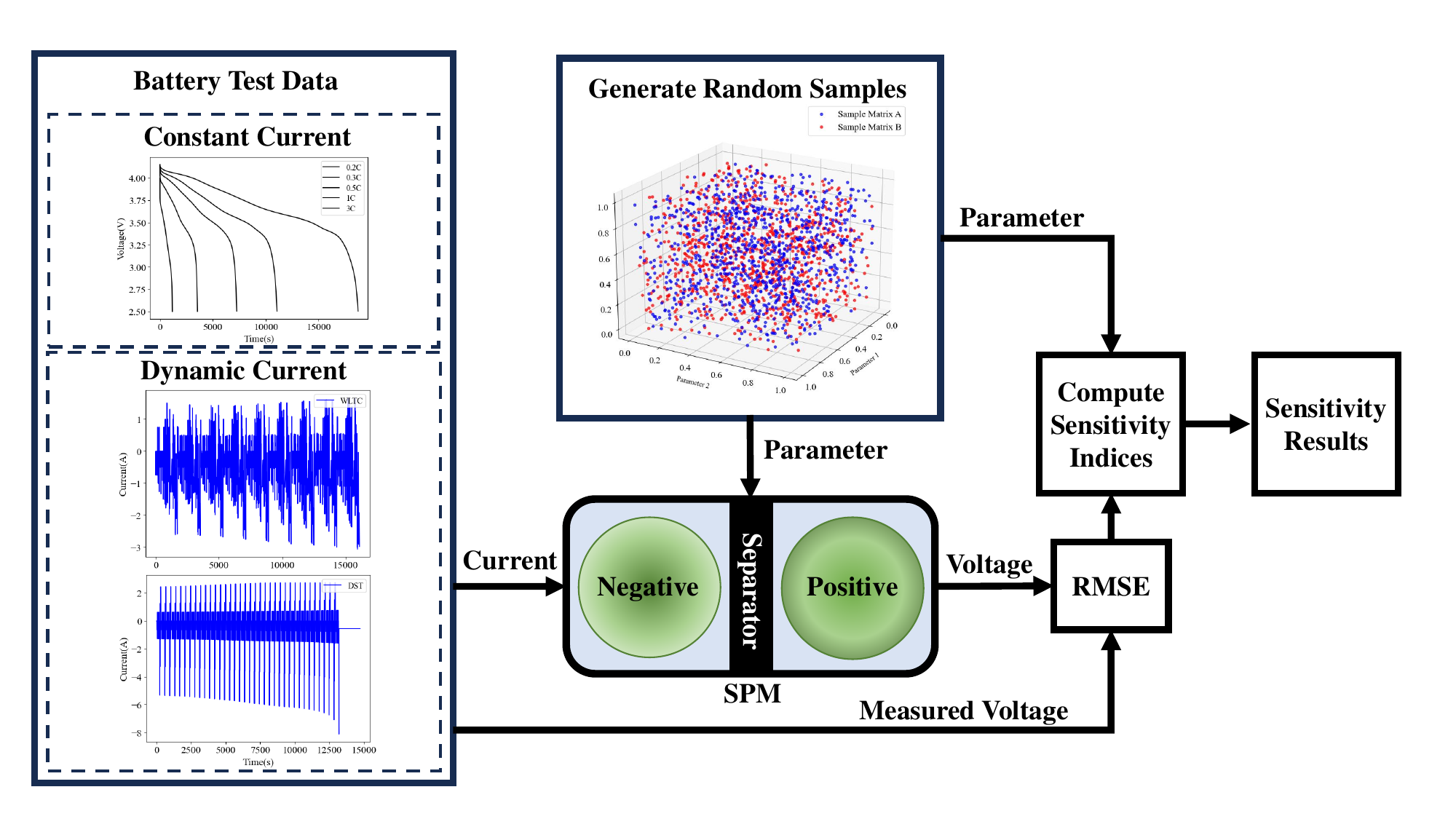}
    \vspace{-0.8cm}
    \caption{Sensitivity analysis using the Sobol method.} 
    \label{fig:Flow}
\end{figure}

In this study, we applied the Sobol method to conduct a global sensitivity analysis on the grouped battery parameters. The Sobol method is a variance-based global sensitivity analysis technique widely used to quantify the contribution of input parameters to the variability of the model output. By decomposing the total variance of the model output, Sobol indices provide detailed insights into the influence of individual input parameters as well as their interactions \cite{campolongo2007effective}. The sensitivity analysis of battery model parameters using the Sobol method is illustrated in Figure \ref{fig:Flow}.

\subsection{Define the Model and Input Parameters}
Define the model:
\begin{equation}
Y = f(P_{\text{grouping}},I,V_{\text{measured}}) ,
\label{eq:46}
\end{equation}
where,  \( Y \) represents the root mean square error (RMSE) between the measured battery voltage and the voltage calculated by the SPM. It is given by:
\begin{equation}
Y = \text{RMSE} = \sqrt{\frac{1}{N} \sum_{i=1}^{N} \left( V_{\text{measured}, i} - V_{\text{SPM}, i} \right)^2},
\label{eq:47}
\end{equation}
where:
\begin{itemize}
    \item \( N \) is the total number of measurements,
    \item \( V_{\text{measured}, i} \) is the \( i \)-th measured battery voltage,
    \item \( V_{\text{SPM}, i} \) is the \( i \)-th voltage calculated by the SPM.
\end{itemize}

This model captures the RMSE between the battery model's voltage and the actual battery voltage under different parameters and current inputs. We use the SPM with grouped parameters after discretization, as described in Section \ref{sec:MDPG}. The parameter vector \( P_{\text{grouping}} \) represents the nine grouped battery parameters from Eq. \eqref{eq:45}. 
The input current to the model is denoted by \( I \). 
To comprehensively evaluate the sensitivity of the battery parameters, we selected both constant current conditions and dynamic current conditions. The constant current conditions include 0.2C, 0.33C, 0.5C, 1C, and 3C discharge scenarios. The dynamic conditions include the DST (Dynamic Stress Test) and WLTC (Worldwide harmonized Light vehicles Test Cycle) profiles. 

\subsection{Generate Random Parameter Samples}
Use Monte Carlo simulation to generate random samples of the input parameters. In this study, a total of 1024 parameter samples were generated. The ranges of these parameters are presented in Table \ref{tab:parameter_ranges}. These ranges are derived from values reported in the literature \cite{fan2020systematic,jin2018parameter,laue2020model, khalik2021parameter,ecker2015parameterization}. However, since the literature provides ranges for the ungrouped parameters, we calculated the corresponding ranges for the grouped parameters used in this study. The values for the positive and negative electrode capacities were determined based on the battery capacity used in our experiments, with a ±20\% variation. The battery capacity utilized in this study is 2.9 Ah. It is important to note that, as shown in Table \ref{tab:parameter_ranges}, there are significant differences in the magnitudes of the various parameters. This disparity could potentially affect the sensitivity analysis. To address this, when calculating the sensitivity, we normalized the range of each parameter to a scale between 0 and 1. This normalization ensures that the sensitivity analysis is not biased by the differing magnitudes of the parameters and allows for a more accurate comparison of their impacts.

\begin{table}[H]
\centering
\caption{Ranges and units of the parameters used in the study.}
\begin{tabular}{lll}
\toprule
\textbf{Parameter} & \textbf{Range} & \textbf{Unit} \\
\midrule
$\alpha_{n}$ & [625, 7692] & %$\text{m}^2/(\text{mol} \cdot \text{m}^{-3})$ \\
$\text{s}$ \\
$\alpha_{p}$ & [1.587, 2500] & %$\text{m}^2/(\text{mol} \cdot \text{m}^{-3})$ \\
$\text{s}$ \\
$Q_n$ & [$2.9 \times 3600 \times 0.8$, $2.9 \times 3600 \times 1.2$] & %$\text{A} \cdot \text{s}$ \\
$\text{C}$ \\
$Q_p$ & [$2.9 \times 3600 \times 0.8$, $2.9 \times 3600 \times 1.2$] & %$\text{A} \cdot \text{s}$ \\
$\text{C}$ \\
$d_{n}$ & [$5.7 \times 10^{-5}$, $7.8 \times 10^{-4}$] & %$\frac{\sqrt{\text{mol} \cdot \text{m}^{-3}}}{\text{m}}$ \\
$\text{s} \cdot \text{mol}^{1/2} \cdot \text{m}^{-5/2}$ \\
$d_p$ & [$7.9 \times 10^{-5}$, $1.0 \times 10^{-3}$] & %$\frac{\sqrt{\text{mol} \cdot \text{m}^{-3}}}{\text{m}}$ \\
$\text{s} \cdot \text{mol}^{1/2} \cdot \text{m}^{-5/2}$ \\
$\text{SOC}_{n}(0)$ & [0.8, 1] & - \\
$\text{SOC}_{p}(0)$ & [0, 0.2] & - \\
$R_0$ & [0, 0.05] & $\Omega$ \\
\bottomrule
\end{tabular}
\label{tab:parameter_ranges}
\end{table}

\subsection{Compute Sensitivity Indices}

The total variance of the model output, \( Var(Y) \), is calculated as:
\begin{equation}
   Var(Y) = \frac{1}{N} \sum_{i=1}^N \left( Y_i - \bar{Y} \right)^2,
\label{eq:48}
\end{equation}
where \( N \) is the number of samples, \( Y_i \) represents the RMSE of the battery model’s output voltage corresponding to each set of parameters, which can be calculated using Eq. \eqref{eq:47} and \( \bar{Y} \) is the mean of the \( Y_i \).

This total variance can be decomposed into contributions from individual input variables and their interactions as follows:
\begin{equation}
Var(Y) = \sum_{n=1}^k Var_n + \sum_{1 \leq n < m \leq k} Var_{nm} + \cdots + Var_{12\ldots k},
\label{eq:49}
\end{equation}
where \( Var_n \) represents the variance contribution of a single input variable \( P_n \), \( Var_{nm} \) represents the contribution from the interaction between \( P_n \) and \( P_m \), and \( Var_{12\ldots k} \) represents the contributions from higher-order interactions.

The total Sobol index, \( S_{Tn} \), includes the contribution of each parameter and all interactions involving that parameter. It is defined as:
\begin{equation}
S_{Tn} = 1 - \frac{Var_{\sim n}}{Var(Y)},
\label{eq:50}
\end{equation}
where \( Var_{\sim n} \) is the variance of the output excluding the contribution of \( P_n \). In this study, we use the total Sobol index \( S_{Tn} \) as the metric for measuring parameter sensitivity.

\section{Results and Discussion} \label{sec:R&D}
In this section, we provide a detailed overview of the simulation process. We also present the results of the sensitivity analysis performed on the grouped parameters, offering insights into their relative impact. Additionally, we compare the results of estimating all grouped parameters with those of estimating only the most sensitive parameters.

\subsection{Simulation Details}
In this study, we used a commercial lithium-ion battery with a nominal capacity of 2.9 Ah. The battery's positive electrode material is Nickel Manganese Cobalt (NMC), while the negative electrode material is graphite. The computational work for battery parameter identification was conducted on a computer equipped with 16 CPUs, and 32 GB of RAM, providing ample processing power for the complex calculations involved. 
 
In this study, we adopted PSO for parameter estimation due to its proven stability and accuracy, as confirmed in our previous work \cite{guo2024efficiency}. Although advanced methods have been recently developed \cite{kim2023strategically}, PSO remains one of the most commonly used techniques in practice \cite{guo2024systematic}. Our goal was to demonstrate that focusing on high-sensitivity parameters can yield comparable accuracy with reduced complexity. So, we chose the widely used PSO method for parameter estimation. Specifically, we utilized the `pyswarms` library in Python to implement the PSO algorithm, ensuring efficient parameter estimation.

We selected the 0.5C constant current discharge condition as the current input for parameter identification. However, the estimated parameters might represent only a locally optimal solution, meaning that while the error is minimized under the 0.5C condition, it could be significantly larger under other conditions. To address this issue,  we first estimated the parameters using the 0.5C current condition. These parameters were then applied to the model. We calculated the voltage RMSE of the model under constant current conditions (0.2C, 0.33C, 0.5C, 1C) as well as under dynamic conditions (DST) using these parameters. In this study, the discharge cut-off voltage for all operating conditions was set at 2.5 V, with the discharge process terminating once this voltage was reached. This RMSE serves as a reference for evaluating the model's overall accuracy across different scenarios. A smaller RMSE indicates better estimated parameters. 

To minimize the impact of stochastic errors introduced by the algorithm, we ran the PSO algorithm 10 times for two of the parameter identification scenarios. In the PSO algorithm, setting the maximum number of iterations too high can lead to excessively long parameter identification times, while setting it too low may cause the algorithm to stop prematurely before converging near the optimal result. In the study by Fan et al. \cite{fan2020systematic}, a maximum of 300 iterations was used, yielding satisfactory results. To prevent premature termination of the PSO algorithm before reaching the optimal solution and to minimize the impact of the maximum iteration limit on the study, this paper opts for a maximum of 500 iterations. The optimization was performed using a swarm of 100 particles to balance exploration capability and computational efficiency. The key hyperparameters were set as follows: the inertia weight (w = 0.9), cognitive coefficient (c1 = 0.5), and social coefficient (c2 = 0.3). These values correspond to the default settings in the 'pyswarms' implementation, which are designed to maintain a balance between individual exploration and collective convergence. It is worth noting that in this study, the same PSO hyperparameters were used for parameter estimation, and the parameter ranges for the battery remained consistent. The parameter ranges specified in Table~\ref{tab:parameter_ranges} were uniformly used as inputs for the PSO method.

\subsection{Sensitivity Analysis Results}
This subsection introduces the use of the Sobol method for conducting sensitivity analysis on the proposed electrochemical model's battery parameters. Figure 2 presents the results of the sensitivity analysis under constant current discharge conditions (0.2C, 0.33C, 0.5C, 1C, 3C). The horizontal axis represents the analyzed battery parameters, while the vertical axis shows the total Sobol sensitivity index. A higher index value indicates greater sensitivity of the parameter to model error; in other words, changes in that parameter have a more significant impact on the voltage error in the battery model output. 

\begin{figure}[H]
    \centering
    \includegraphics[width=\textwidth,trim={0.3cm 0.3cm 0.3cm 0.3cm},clip]{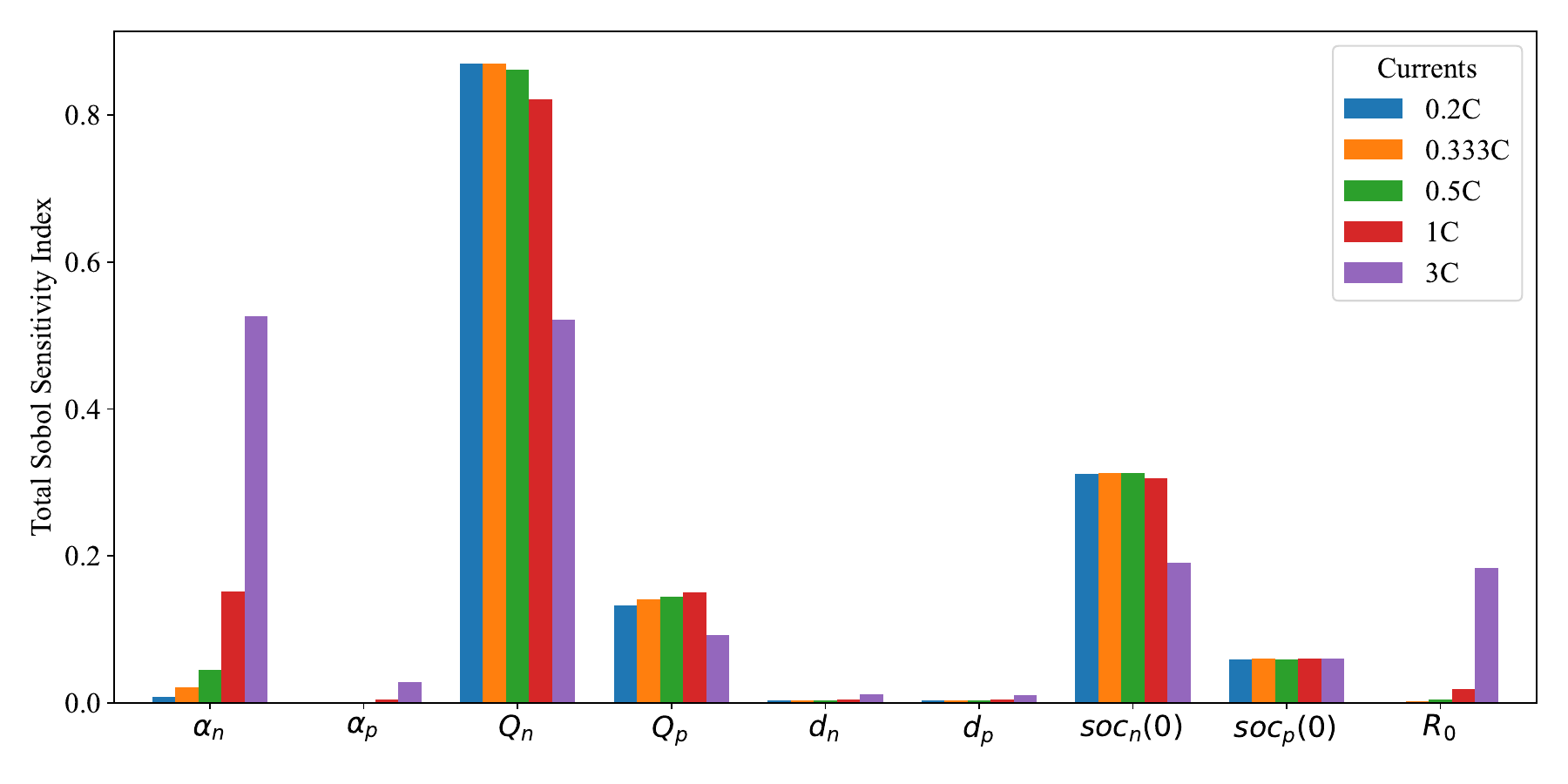}
    \vspace{-0.8cm}
    \caption{Sensitivity analysis results for constant current.} 
    \label{fig:constant current}
\end{figure}

As shown in Figure 2, it is evident that the parameters \( \alpha_p \), \( d_n \), and \( d_p \) exhibit consistently low sensitivity across all constant current discharge conditions. Since these parameters have a minimal impact on the overall model accuracy, they are not estimated during the parameter estimation process. Theoretically, this approach can simplify the model without compromising its accuracy. Additionally, the sensitivity values for \(Q_n\) and \({\rm SOC}_n\) are higher compared to other parameters. This indicates that in this electrochemical model, the battery capacity and the initial SOC of the negative electrode have the most significant impact on model error. If designing a battery SOC estimation algorithm based on this model, the SOC of the negative electrode is more suitable than the positive electrode as an indicator for SOC estimation. 

When comparing different constant current discharge conditions, it is evident that the sensitivity of parameters \( \alpha_n \) and \( R_0 \) increases with rising current rates, with \( R_0 \) showing a particularly pronounced growth in sensitivity at higher currents. This significant increase is attributed to the fact that, as shown in Eq. \eqref{eq:4}, the current and internal resistance \( R_0 \) are multiplied together as part of the model output voltage. Therefore, as the current increases, the influence of \( R_0 \) on the overall model voltage output becomes substantially more significant. Moreover, the rising sensitivity of \( R_0 \) and \( \alpha_n \) underscores the need to pay particular attention to these parameters under high-rate discharge conditions. This is crucial for practical applications, as it has significant implications for BMS to ensure both safety and performance under high-load scenarios. The sensitivity indices of parameters \( Q_n \), \( Q_p \), and \( {\rm SOC}_{n}(0) \) remain relatively stable across different current rates, with only a noticeable decrease in sensitivity under the 3C constant current discharge condition. This stability suggests that these parameters have a consistent influence on the model output across a range of operating conditions. Additionally, the sensitivity of \( {\rm SOC}_{p}(0) \) remains nearly constant across all constant current discharge conditions, further indicating that the impact of this parameter on the model's output is uniform across various operational scenarios.

\begin{figure}[H]
    \centering
    \includegraphics[width=\textwidth,trim={0.3cm 0.3cm 0.3cm 0.3cm},clip]{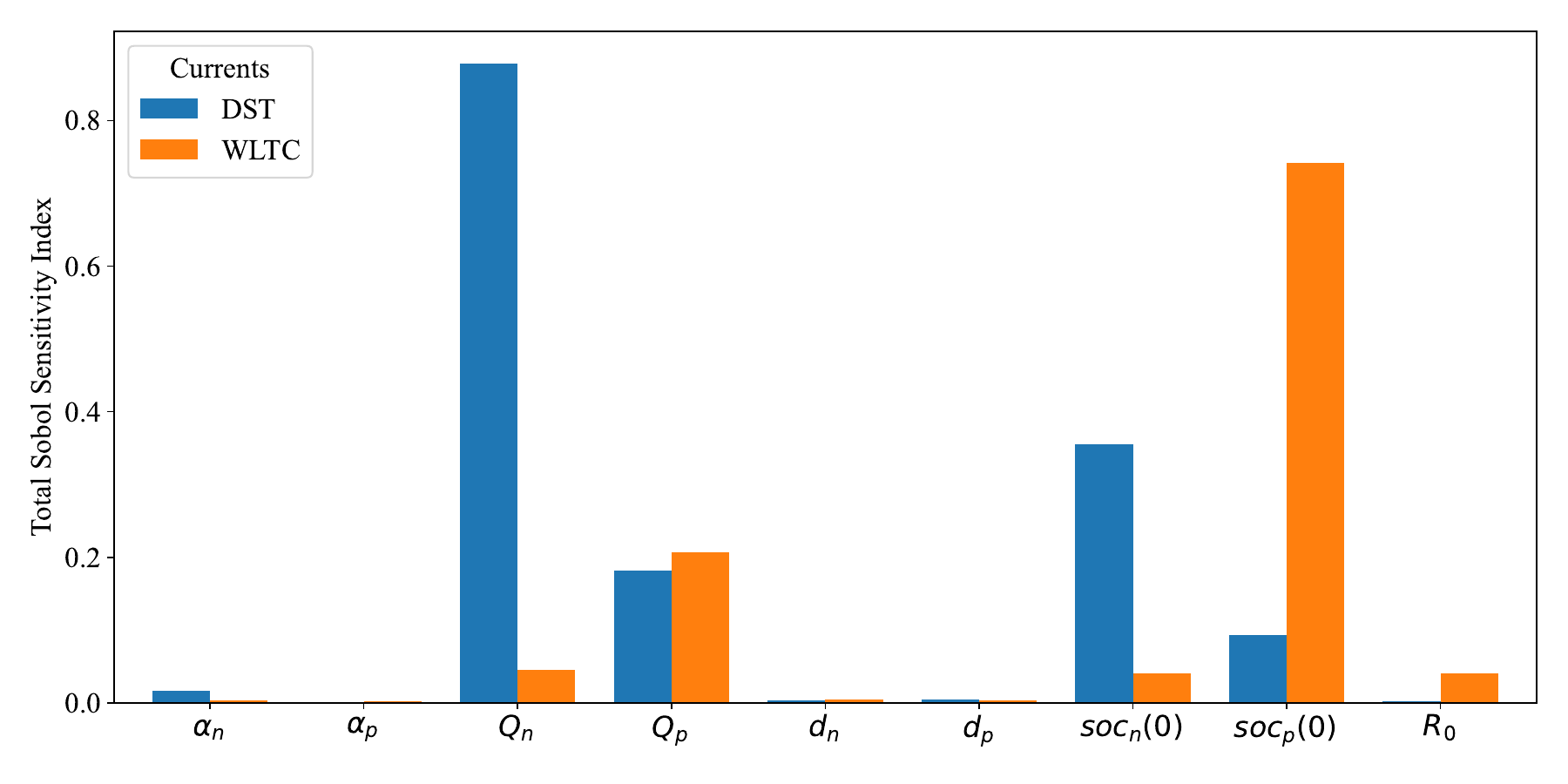}
    \vspace{-0.8cm}
    \caption{Sensitivity analysis results for dynamic current.} 
    \label{fig:dynamic current}
\end{figure}

Figure~\ref{fig:dynamic current} presents the sensitivity analysis results for various parameters when the dynamic conditions DST and WLTC are used as model inputs. The results indicate that the parameters \( \alpha_p \), \( d_n \), and \( d_p \) exhibit the lowest sensitivity under dynamic conditions, with values close to zero, which is consistent with the results from constant current discharge conditions. This suggests that these three parameters do not require estimation during the parameter estimation process. 

The parameters \( Q_n \), \( {\rm SOC}_n \), \( Q_p \), and \( {\rm SOC}_p \) demonstrate the highest sensitivity. Notably, the negative electrode parameters \( Q_n \) and \( {\rm SOC}_n \) show high sensitivity under the DST condition, aligning with the results from constant current discharge. However, under the WLTC condition, the sensitivity of the positive electrode parameters \( Q_p \) and \( {\rm SOC}_p \) surpasses that of the negative electrode parameters \( Q_n \) and \( {\rm SOC}_n \). This indicates that the model responds differently to varying types of dynamic current inputs. Compared to the pulse current in DST, the WLTC current exhibits more intense fluctuations and a greater degree of dynamic characteristics. The higher sensitivity of \( Q_p \) and \( {\rm SOC}_p \) under WLTC conditions suggests that these positive electrode parameters are more influential when the current undergoes rapid changes.

In light of these findings, it is crucial to prioritize the calibration of the parameter \( Q_n \) under DST conditions, while the initial value setting of \( {\rm SOC}_p(0) \) becomes particularly important under WLTC conditions. Additionally, the parameter \( \alpha_n \) shows higher sensitivity in DST compared to WLTC, with its sensitivity under WLTC approaching zero, indicating that this parameter is less responsive to rapidly changing currents. In contrast, the parameter \( R_0 \) behaves oppositely, with higher sensitivity under WLTC and sensitivity nearing zero under DST conditions. Based on the comprehensive sensitivity analysis of battery parameters under both constant current and dynamic conditions (in Figure~\ref{fig:constant current} and Figure~\ref{fig:dynamic current} ), we can conclude that the parameters \( \alpha_p \), \( d_n \), and \( d_p \) exhibit consistently low sensitivity across both types of conditions. And they are significantly lower than that of other parameters and approaches zero. Therefore, these parameters can be excluded from the estimation process. When estimating other parameters, any arbitrary value within their respective ranges can be assigned to these three parameters.This allows us to reduce the number of parameters that need to be estimated to six. The high-sensitivity parameters that require estimation are as follows:

\begin{equation}
\begin{aligned}
P_{\text{high-sensitivity}} = \big[ \alpha_{n}, Q_n, Q_p, {\rm SOC}_{n}(0), {\rm SOC}_{p}(0),R_0 \big]
\end{aligned}
\label{eq:51}
\end{equation}

\subsection{Comparison of Different Numbers of Estimated Parameters}

Since the true battery parameters are typically unavailable, the accuracy of parameter estimation is generally evaluated by comparing the error between the output voltage of the battery model and the actual measured voltage. This approach ensures that the estimated parameters can effectively replicate the real battery behavior, with smaller voltage errors indicating higher parameter estimation accuracy. The experimental procedure for parameter estimation can be referenced in Section 5.1. To eliminate the influence of random errors, we ran the PSO parameter estimation algorithm 10 times for each scenario. Consequently, we obtained 10 sets of estimated battery parameters for each case. 

It is worth noting that when estimating only the highly sensitive parameters (6 parameters), the remaining three parameters ($\alpha_{p}$, $d_n$, and $d_p$) are assigned arbitrary values within their respective parameter boundaries (as shown in Table~\ref{tab:parameter_ranges}). In this study, we set these values to 1250, 0.00005, and 0.0005, respectively. Since the previous section has already demonstrated that these parameters have minimal impact on the voltage output error of the battery model, they can be freely assigned within their specified ranges.

Table~\ref{tab:rmse_results1} presents the average RMSE results across all operating conditions for 10 sets of parameters obtained separately from the estimation of all grouped parameters (9 parameters) and high-sensitivity parameters (6 parameters). Additionally, the average RMSE, minimum RMSE, and maximum RMSE for the 10 sets of high-sensitivity parameters (6 parameters) are all lower than those for the 10 sets of all grouped parameters (9 parameters), indicating improved accuracy when focusing on high-sensitivity parameters. Moreover, the standard deviations of both sets of results are low, demonstrating the stability of the PSO-based parameter estimation. This suggests that estimating only the high-sensitivity parameters can achieve accuracy comparable to estimating all parameters, reducing computational complexity without sacrificing precision. 

\begin{table}[h]
    \centering
    \caption{Comparison of Average Errors for 10 Parameter Sets Obtained Using PSO: High-Sensitivity vs. All-Parameter Estimation Across All Operating Conditions}
    \begin{tabular}{lcc}
        \toprule
        \textbf{Statistic} & \textbf{9 Parameters} & \textbf{6 Parameters} \\
        \midrule
        Mean (V) & 0.0366 & 0.0340 \\
        Standard Deviation (V) & 0.0059 & 0.0061 \\
        Maximum Value (V) & 0.0503 & 0.0458 \\
        Minimum Value (V) & 0.0299 & 0.0281 \\
        \bottomrule
    \end{tabular}
    \label{tab:rmse_results1}
\end{table}

Table~\ref{tab:rmse_results2} presents the best parameter sets obtained from the parameter estimation conducted for both the high-sensitivity parameters and the complete set of parameters. It can be observed that the RMSE values under various operating conditions for the high-sensitivity parameters are consistently lower than those for the full parameter estimation. This further demonstrates that estimating only the high-sensitivity parameters is sufficient to ensure the accuracy of the battery model. Additionally, it is noteworthy that, in both cases, the error at the 0.5C condition is the lowest. This is attributable to the fact that the parameter estimation was performed using the 0.5C operating condition, which naturally yields minimal error under these conditions.

\begin{table}[h]
\centering
\caption{RMSE under Various Operating Conditions using PSO Best Parameter Results}
\label{tab:rmse_results2}
\begin{tabular}{lcccccc}
\toprule
 & \multicolumn{1}{c}{0.2C (V)} & \multicolumn{1}{c}{0.33C (V)} & \multicolumn{1}{c}{0.5C (V)} & \multicolumn{1}{c}{1C (V)} & \multicolumn{1}{c}{DST (V)} & \multicolumn{1}{c}{Mean (V)} \\
\midrule
6 Parameters & 0.0370 & 0.0199 & 0.0158 & 0.0262 & 0.0415 & 0.0281 \\
9 Parameters & 0.0404 & 0.0213 & 0.0163 & 0.0289 & 0.0426 & 0.0299 \\
\bottomrule
\end{tabular}
\end{table}

\begin{figure}[H]
    \centering
    \includegraphics[width=\textwidth,trim={0.3cm 0.3cm 0.3cm 0.3cm},clip]{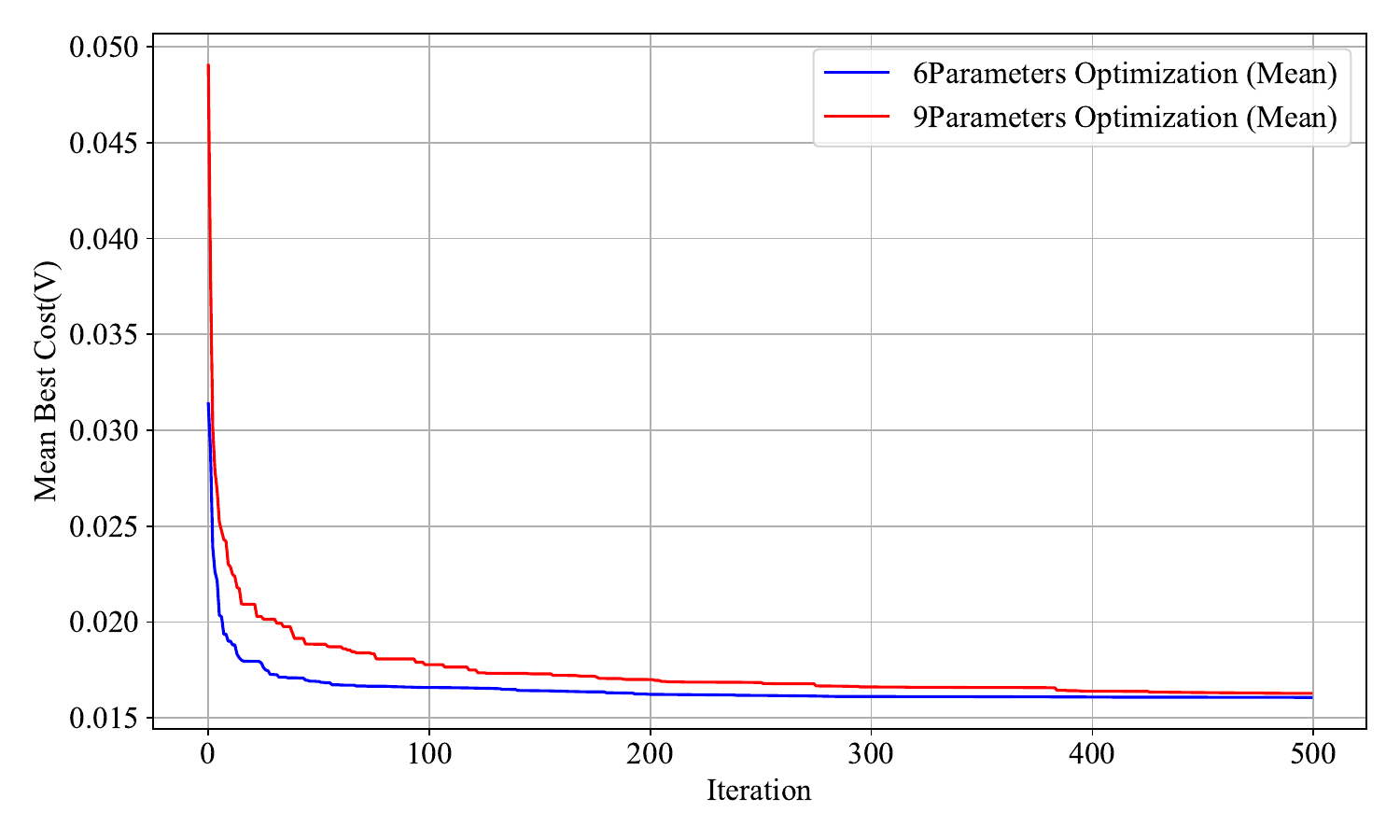}
    \vspace{-0.8cm}
    \caption{PSO convergence comparison.} 
    \label{fig:PSO}
\end{figure}

Figure \ref{fig:PSO} illustrates the results of running the PSO process. The horizontal axis represents the number of PSO algorithm iterations, with a maximum of 500 iterations. The vertical axis shows the average best result for each iteration. It is evident that the blue line, which represents the case where only high-sensitivity parameters (6 parameters) are considered, consistently lies below the red line, which represents the case where all grouped parameters (9 parameters) are estimated. This demonstrates a faster convergence rate when only the high-sensitivity parameters are estimated. To further illustrate the computational advantage of estimating 6 parameters over 9 parameters, our analysis shows that estimating 6 parameters requires only 193 iterations to achieve a similar accuracy level (0.016249 RMSE) as 9 parameters after 500 iterations (0.016276 RMSE). This result demonstrates that reducing the number of estimated parameters can significantly improve computational efficiency, achieving an approximately 2.6 times speedup while maintaining comparable model accuracy. The reason for this faster convergence is that by focusing solely on the high-sensitivity parameters, the number of parameters to be estimated is reduced, allowing the PSO algorithm to explore a smaller parameter space and find the global optimum more quickly. This also confirms that considering only the high-sensitivity parameters can improve the convergence speed of the parameter estimation process. It can be observed that, in the final stages, the best cost values for both methods level off and no longer decrease rapidly. This indicates that the PSO algorithm has likely found a solution close to the global optimum.

In summary, the results indicate that estimating only the high-sensitivity parameters can achieve the same level of model error as estimating all grouped parameters. Additionally, focusing on high-sensitivity parameters leads to a faster convergence rate during the parameter estimation process.

\section{Conclusion}
\label{sec:Conclu}

In this study, we proposed a control-oriented SPM based on a parabolic approximation method for battery state estimation and control. By reformulating the model to use the average lithium concentration and surface lithium concentration, we aligned the model with state estimation and control requirements while reducing parameter complexity. Parameter grouping, followed by a comprehensive global sensitivity analysis using the Sobol method under various constant and dynamic current profiles, further refined the selection of model parameters requiring identification. As a result, the original 17 parameters were reduced to 9 grouped parameters and then further condensed to 6 highly sensitive parameters. The PSO method confirmed that estimating only these critical parameters achieves accuracy comparable to full-parameter estimation, with improved computational efficiency. These insights indicate that sensitivity analysis should be tailored to the specific operational conditions of the battery in practical applications, which can guide the design of BMS.

It should be noted that the sensitivity analysis results in this study are specific to the input conditions and battery materials used. Further research is needed to validate these findings across different battery types and operational conditions. In practical applications, the methods proposed in this study can be adapted to the specific batteries and their operating conditions.

% \section*{Acknowledgements}
% We would like to thank [Name or Institution] for [specific help or funding].

\section*{Conflict of Interest}
The authors declare that they have no conflict of interest.

%% The Appendices part is started with the command \appendix;
%% appendix sections are then done as normal sections

\appendix
\label{sec:appendix}
\section{OCP Formulation for the SPM}

In this appendix, we provide the analytical expressions for the OCP used in the SPM of this work. The OCP for the positive electrode is represented by a seventh-order polynomial, whereas the OCP for the negative electrode is described by a combination of exponential and hyperbolic tangent functions. The coefficients in these expressions have been determined through experimental data fitting and are listed explicitly below.

\subsection{Positive Electrode}
The open-circuit potential of the positive electrode, denoted as $OCP_p(x)$, where $x=\tilde{c}_{ss,p}(t)$ represents the normalized surface concentration, is given by
\begin{equation}
\begin{aligned}
OCP_p(x) =\; &4.26541327 - 1.74561881\, x + 12.91342685\, x^2 - 71.23523821\, x^3 \\
&+ 182.39441925\, x^4 - 237.12698576\, x^5 + 153.41883911\, x^6\\
&- 39.38243997\, x^7.
\end{aligned}
\end{equation}

\subsection{Negative Electrode}
The open-circuit potential of the negative electrode, denoted as $OCP_n(x)$, with $x=\tilde{c}_{ss,n}(t)$ representing the corresponding normalized surface concentration, is formulated as
\begin{equation}
\begin{aligned}
OCP_n(x) =\; &50.9268468 - 1.01981453\, \exp\!\left(0.00300736348\, x\right) \\
&- 97.3001503\, \tanh\!\left(\frac{x + 0.133226619}{0.0438465709}\right) \\
&- 0.0188982006\, \tanh\!\left(\frac{x - 0.511704304}{0.0237662771}\right) \\
&- 0.0437651724\, \tanh\!\left(\frac{x - 0.174973549}{0.0540321443}\right) \\
&- 47.5291964\, \tanh\!\left(\frac{x - 1.18796536}{0.0516560110}\right).
\end{aligned}
\end{equation}

%% If you have bibdatabase file and want bibtex to generate the
%% bibitems, please use
%%
 \bibliographystyle{elsarticle-num} 
 \bibliography{cas-refs}

%% else use the following coding to input the bibitems directly in the
%% TeX file.

% \begin{thebibliography}{00}

% %% \bibitem{label}
% %% Text of bibliographic item

% \bibitem{}

% \end{thebibliography}
\end{document}